# First Experiences with LHC Grid Computing and Distributed Analysis

I. Fisk on behalf of the CMS collaboration
*FNAL, Batavia, IL 60510, USA*

In this presentation the experiences of the LHC experiments using grid computing were presented with a focus on experience with distributed analysis. After many years of development, preparation, exercises, and validation the LHC (Large Hadron Collider) experiments are in operations. The computing infrastructure has been heavily utilized in the first 6 months of data collection. The general experience of exploiting the grid infrastructure for organized processing and preparation is described, as well as the successes employing the infrastructure for distributed analysis. At the end the expected evolution and future plans are outlined.

## 1. INTRODUCTION

Computing for the LHC experiments has grown up with the development of grid models. The original distributed computing models were based, to greater and lesser extents, on the MONARC report [1]. MONARC introduced in 2001 the concept of a multi-tiered computing infrastructure with defined functionality and connectivity at each tier. The Worldwide LHC Computing Grid (WLCG) was approved by the CERN council in 2001 and was tasked with providing services, infrastructure, and organizing computing facilities for the four LHC experiments [2,3]. In order to make the globally distributed set of computing facilities in multiple tiers function as a coherent computing system the WLCG collaborated with grid projects in Europe and the U.S. on interfaces and services to enable grid computing.

Today the WLCG is comprised of more than 140 sites on 5 continents. The combined resources execute more than 1 million job submissions per day, have access to 150k processor cores and more than 50 peta-bytes of disk. This paper describes the initial experiences of the four LHC experiments (ALICE, ATLAS, CMS and LHCb) using the grid infrastructure during the first 6 months of running.

## 2. ARCHITECTURE AND ACTIVITIES

Most of the LHC computing models are based on 3 central computing tier levels. The Tier-0 facility at CERN performs the prompt reconstruction of the data, low latency workflows for calibration and detector commissioning, archival storage of the data, and distribution of samples to the Tier-1 centers. The Tier-1 centers are connected to CERN via an optical private network (OPN). The Tier-1s are responsible for reprocessing of the data samples, custodial storage of data and simulation, and serving the data to the Tier-2s. The Tier-2s for most of the LHC experiments are the primary resources used for analysis and for all the experiments are the largest sources of simulated event production. It varies by experiment and year, but CERN is roughly 20% of the total processing capacity with 40% at both the Tier-1 and Tier-2 centers. The LHC experiments, even in the first few months of data collection, were performing the predicted workflows at the intended facility. There had been concerns about the challenges of commissioning a distributed computing system at the same time as commissioning the new detectors with data, but generally the experience using the grid sites for their intended purpose has been good and the operations programs of all the experiments have been successful in processing, distributing, and analyzing the data samples.

## 2.1. Commissioning

The smoothness of the LHC Computing start can be partially credited to the long commissioning program. All four experiments have been operating regular data and service challenges at increasing scale since 2004. The goal of these tests was to exercise the computing systems and determine what components were operating and functioning at the expected scale, and which elements needed more design and development. Not all the challenges were equally successful in terms of validating components, but all were useful to direct plans and effort. The final two challenges were the Common Computing Readiness Challenge[4] (CCRC08) that demonstrated the full functionality of the computing models operating simultaneously to look for interference effects, and the Scale Test of the Experiment Program (STEP09) that aimed to demonstrate simultaneous interactions at the full scale expected at the start of the experiment. The regular testing with simulated event samples during the years of preparation ensured that the computing systems functioned under real data conditions during the first year.

## 2.2. Testing and Monitoring

The global distribution of the sites and the complexity of the services drives a need for continuous monitoring of the sites for availability. Since 2006 WLCG has been monitoring the availability and reliability of the sites to check their readiness. The Tier-1 availability is considered particularly important because those site store the second custodial copy of the raw data, and the site availability level is specified by Memorandums of Understanding (MOU) [5]. Since the tests began the availability has steadily improved for the Tier-1 and Tier-2 sites and most Tier-1s now achieve the goals outlined in the MOU. The average readiness of the best 8 sites began at around 85% and over 4 years have improved to the high nineties. Experiment specific workflows were added more recently to allow a more detailed view of the site performance. The advent of collision data in the LHC program is visible in the readiness views as the sites became more conscientious. The careful monitoring of the sites and the operations effort has allowed the LHC experiments to rely on a much larger collection of resources to complete the computing activities even in the first year of data.

## 3. OPERATIONS

The first year of data collection presents unique challenges because the luminosity increases exponentially. During the first several months the accumulated luminosity could be eclipsed by a good weekend, or at the beginning even a good fill. By the time 1 inverse pico-barn had been collected the instantaneous luminosity increased by 3 orders of magnitude from the level at the beginning of run. During these rapidly changing conditions it is important to maintain stability in the computing systems, because even a short failure could result in a big percentage loss of the sample. The LHC experiments have been able to utilize the computing facilities to promptly reconstruct the incoming events and archive them at the Tier-0 at CERN. From CERN the data has been successfully transferred to remote Tier-1 computing facilities to manage a second archival copy and to provide computing resources for reprocessing. Analysis samples are transferred to Tier-2s. Simulation samples are generated at both Tier-1s and Tier-2s and transferred to Tier-1s for permanent storage.

### 3.1. Data Distribution

After substantial commissioning effort all four LHC experiments are able to efficiently move data between the computing tiers. The traffic from CERN to Tier-1s is carried on the dedicated OPN, where all connections to Tier-1s have reached 10Gb/s and back-up channels exist to handle temporary service interruptions. During 2010 running at around the ICHEP conference the total rate from CERN had daily averages over 2GB/s (16Gb/s) per second from the combination of the 4 experiments. ATLAS had the highest volume of raw data being distributed and made up about 60% of the total. Figure 1 shows the transfers from CERN during this period. The rate from CERN during ICHEP was similar to the rates achieved in the last service challenge of 2009 in total export rate and experimental contribution. This indicates the final full-scale service challenge was representative of the first year of running.

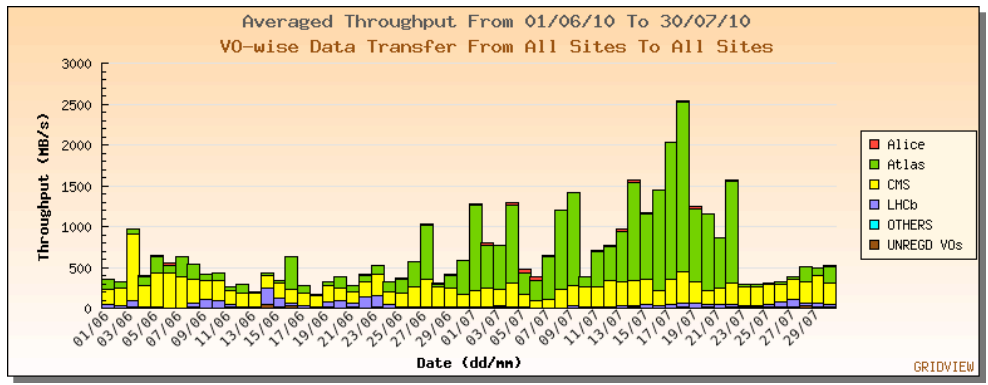

Figure 1: Transfer rates from CERN to Tier-1s during the preparation for the summer conferences.

The OPN is also used by the experiments to replicate data between Tier-1s. The total available bandwidth including all the connections adds to roughly 120Gb/s. During peak periods of synchronizing recently reprocessed data across Tier-1s rates of 70Gb/s have been seen across the network. This is driven by replicating reconstructed data across all the Tier-1s by ATLAS, but spikes of 40Gb/s are reasonably routine under normal experiment activity.

Once data has been successfully transferred to Tier-1 sites, in the case of most of the LHC experiments the Tier-1 sites are responsible for serving datasets to Tier-2 sites for further analysis by users. The networking challenge to serve data can be more than the load caused by ingesting data. There are more Tier-2 sites than there are Tier-1s by roughly a factor of 5, so there are more potential destinations. Additionally, the rate from CERN to Tier-1s is controlled by the trigger rate from the experiments, and therefore roughly predictable. The rate to Tier-2s from Tier-1s is controlled by the activity level of the users and can vary rapidly as interest in datasets evolve. The two experiments with the broadest data distribution at the Tier-2 level are ALICE and CMS. ALICE has adopted an xrootd[6] based model for data distribution where an application can receive the data from the closest available copy. CMS has adopted a full mesh model where complete data samples are replicated on demand, but can be sent from any Tier-1 to any Tier-2. A growing number of Tier-2 to Tier-2 transfers are now used in production. The transfer rates achieved within a region are several hundred mega-bytes per second. Between regions, including transatlantic, 100 mega-bytes per second between regions is becoming more common.

## 3.2. Data Reprocessing

Once data has been successfully transferred to Tier-1 sites and stored, the Tier-1 centers are responsible for reprocessing the samples. Reprocessing has been done frequently during the first year as code and conditions information has improved. Submitting reprocessing requests involves a large number of submissions through the grid computing elements and then the successful interactions with the grid storage elements. This workflow has been successfully demonstrated by all the experiments. The total volume of data at the LHC in the first year is not as large as what would be expected in a nominal data taking year, and the experiments are able to reprocess the complete dataset quickly. Using all 10 of their Tier-1 sites ATLAS was able to process the data in the late spring of 2010 in 12 days with each facility taking the commensurate share of the raw data. The fastest site to complete their share was 3 days and the second slowest was 1 week. ALICE and CMS have demonstrated performance in reprocessing the collected data with both experiments demonstrating the use of all 7 of their Tier-1 sites with a total of more than 10k running jobs simultaneously.

## 4. ANALYSIS

One of the biggest accomplishments of the experiments in collaboration with the grid projects is the successful transition of the analysis workflow to the distributed infrastructure. A general concern during the commissioning phase was whether users would adopt the tools and what challenges would be faced trying to move users from performing analysis at CERN. Considerable investments were made by all the experiments in insulating the user from the intrinsic complexity of the distributed grid environment and attempting to make a complicated chain of communication look like a simple batch queue submission. When an analysis job is submitted to the grid the user environment is packaged, so it is available at the remote location; the datasets requested are resolved and located on remote resources; the individual job sections are submitted through the computing element; and the results are returned to the users. There is a lot of custom development in the individual experiment systems to support experiment specific applications and concepts in data management, but the basic functionality is similar.

The most impressive aspect of the analysis frameworks of the experiment is not the level of transparency achieved, but the level of adoption achieved. The smaller collaborations of ALICE and LHCb both have more than 200 active users per month submitting jobs to the grid. ATLAS and CMS have 1000 and 800 individual submitters per month respectively. The number of individuals submitting per week in the CMS experiment is shown as an example in Figure 2. The peak on the left of the curve is an analysis preparation activity, while the steady increase on the right is the first months of data and the lead up to the ICHEP conference. The collaborations are much larger than the number of submitters, but the majority of active analysis users are performing their analyses with the help of grid resources. The analysis submissions are a sizeable fraction of the total jobs on the grid. CMS currently has approximately 100k grid submissions per day from analysis, which is roughly what was predicted in the computing model. All the experiments have been able to make good use of the many distributed Tier-2 computing facilities.

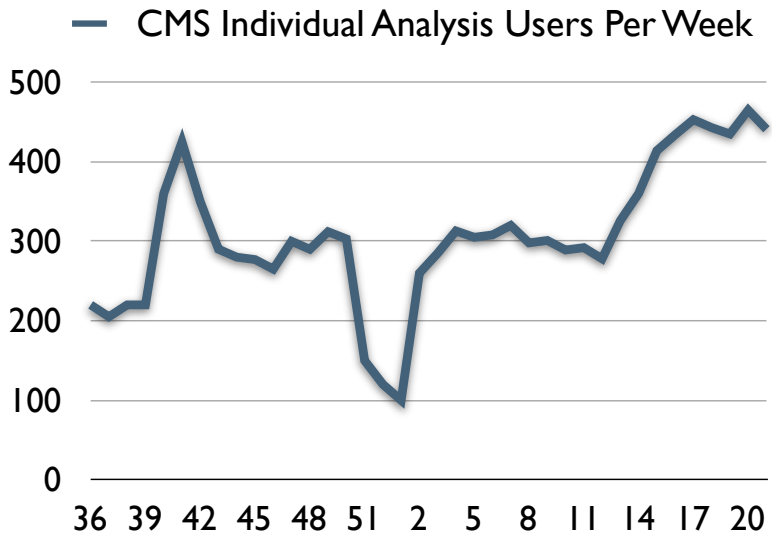

Figure 2: CMS individual analysis users per week from mid 2009 to mid 2010

## 5. LOOKING FORWARD

The LHC computing models have been under development, validation, and testing for almost 10 years.  Over this time there has been a continuous evolution in the underlying technologies and protocols used.  This trend is expected to continue with several new technology changes expected on the horizon that will need to be addressed.   The continued push to more cores per physics processor is expected to continue toward very large numbers of cores.  The number of cores per processor has almost doubled yearly with 6 and 8 core processors in production.    The continued increase in the number of running processes needed to fully exploit the processing capabilities is driving the experiments to examine the workflow tools and the design of the applications to ensure reasonably processing efficiency.  While the experiments are probably some time away from true multi-threaded, multi core applications, utilization of multiple cores at the workflow level is probably closer.   Another interesting technology change being investigated is the migration to more virtualized systems and cloud computing.   The possibility of adding resources in commercial clouds as needed is being investigated by several of the LHC experiments, but successful working in computing clouds implies challenges of data management and data access that must be overcome.

In addition to technology evolution, the experiments are actively looking for ways to improve operational efficiency and make the best use of the computing resources as the volume of data increases.  The current techniques and infrastructure have been successfully deployed for data collection, reconstruction, and analysis.   The experiments are able to fully utilize the available computing resources and physics results are being published, but looking forward there may be efficiency gains in rethinking how data access and data management is handled.  With the possible exception of ALICE, the LHC experiments use very deterministic models.   The computing jobs are sent to data that was previously replicated.   The static placement of data ensures that samples are accessible to users, but needs substantial disk resources to host samples that may not be frequently accessed.  A lot of the anticipated transparency across computing facilities and the ability to balance load and use has not been fully realized.   The distributed computing system is a cluster of independent centers with defined and managed interfaces, which is functional but additional flexibility could be achieved by reducing some of the barriers between centers. Some of the assumptions of the MONARC models are being reconsidered in light of improved networking and evolutions of technology. LHC experiments are looking at more dynamic methods for accessing data over the wide area, accessing data through true caches, and transferring data on demand. These changes have the potential for making more flexible use of the available computing centers and providing functionality to adopt cloud computing models.

## 6. OUTLOOK

The grid infrastructure is working for LHC physics: data is reprocessed, stored and delivered to analysis users. The challenging task of deploying a distributed computing system in operations while commissioning a new detector was a success, due in part to the detailed testing and validation program between the experiments and the WLCG. In the first year, the integrated volume of data is still relatively small and subsequently the experiments have a lot of computing resources. There will be an interesting transition to a more resource-constrained environment coming. Even in the first year the user activity level and enthusiasm are high.